\documentclass[useAMS,usetimes,usemathptm,usenatbib,usegraphicx]{mn2e}

\title[Gravitational lensing in a non-uniform plasma]{Gravitational lensing in a non-uniform plasma}
\author[G. S. Bisnovatyi-Kogan and O. Yu. Tsupko]{G. S. Bisnovatyi-Kogan$^{1,2}$\thanks{E-mail:
gkogan@iki.rssi.ru (GSBK); tsupko@iki.rssi.ru (OYuT)} and O. Yu. Tsupko$^{1,2}$\footnotemark[1]\\
$^{1}$Space Research Institute of Russian Academy of Sciences, Profsoyuznaya 84/32, Moscow 117997\\
$^{2}$National Research Nuclear University MEPhI, Kashirskoe Shosse 31, Moscow 115409, Russia}

\newcommand{\arctanh}{\mathop{\rm arctanh}\nolimits}

\begin{document}



\maketitle

\label{firstpage}

\begin{abstract}
We develop a model of gravitational lensing in a non-uniform plasma. When a gravitating body is surrounded by
a plasma, the lensing angle depends on the frequency of the electromagnetic wave, due to dispersion
properties of plasma, in presence of a plasma inhomogeneity, and of a gravity. The second effect leads, even
in a uniform plasma, to a difference of the gravitational photon deflection angle from the vacuum case, and
to its dependence on the photon frequency. We take into account both effects, and derive the expression for
the lensing angle in the case of a strongly nonuniform plasma in presence of the gravitation. Dependence of
the lensing angle on the photon frequency in a homogeneous plasma resembles the properties of a refractive
prism spectrometer, which strongest action is for very long radiowaves. We discuss the observational
appearances of this effect for the gravitational lens with a Schwarzschild metric, surrounded by a uniform
plasma. We obtain formulae for the lensing angle and the magnification factors in this case and discuss a
possibility of observation of this effect by the planned VLBI space project Radioastron. We also consider
models with a nonuniform plasma distribution. For different gravitational lens models we compare the
corrections to the vacuum lensing due to the gravitational effect in plasma, and due to the plasma
inhomogeneity. We have shown that the gravitational effect could be detected in the case of a hot gas in the
gravitational field of a galaxy cluster.
\end{abstract}


\section{Introduction}

The photon deflection angle in vacuum, in the Schwarzschild metric with a given mass $M$, is determined, for
small deflection angles $\hat\alpha\ll1$, by a formula
\begin{equation}
\label{Einst-angle} \hat{\alpha} = \frac{4GM}{c^2 b} = \frac{2 R_S}{b} ,
\end{equation}
where $b$ is the impact parameter, and $b\gg R_S$,  $R_S=2GM/c^2$ is the Schwarzschild radius \citep*{LL2,
MTW, GL}. When the impact parameter is close to its critical value, corresponding to the capture of the
photon by the black hole, the expression for the deflection angle is more complicated \citep{Darwin, MTW,
Virbh, Bozza2001, BK&TsStrong}.

It is interesting to consider the gravitational lensing in a plasma, because in space the light rays mostly
propagate through this medium. In an inhomogeneous plasma photons move along the curved trajectories, because
a plasma is a dispersive medium with a permittivity tensor depending on its density \citep{LL8}. In the
dispersive nonuniform medium a photon trajectory depends also on the photon frequency, and this effect has no
relation to the gravity. The photon deflection in a non-homogeneous plasma, in presence of gravity, has been
considered by \citet{B-Minakov}, \citet*{Muhl1970}, \citet{zadachnik}. The consideration was done in a linear
approximation, when the two effects: the vacuum deflection due to the gravitation, and the deflection due to
the non-homogeneity of the medium, have been considered separately. The first effect is achromatic, the
second one depends on the photon frequency if the medium is dispersive, but equals to zero if the medium is
homogeneous.

A general theory of the geometrical optics in a curved space-time, in arbitrary medium, is presented in the
book of \citet{Synge}. On the basis of his general approach we have developed the model of gravitational
lensing in plasma. In our previous work \citep{BK&Ts2009} we have shown, that in a nonlinear approach a new
effect appears. Even in the homogeneous plasma, the gravitational deflection angle really depends on the
frequency of the photon, and is different from the vacuum case. We have derived the expression for the
deflection angle of the photon in a weak gravitational field, in a weakly inhomogeneous plasma.

In this work we use more general approach and derive the deflection angle for the photon moving in a weak
gravitational field, in the Schwarzschild metric, in the arbitrary inhomogeneous plasma. Such approach is
more appropriate for the propagation in a cosmic plasma, because plasma density changes significantly, from
the density of the interstellar medium to the density of a black hole neighborhood. We consider here only the
situation, when the whole deflection angle, from the combined plasma and gravity effects, remains small. In
the section 2 we derive a general expression for the deflection angle in the inhomogeneous plasma, in the
curved space-time. In the section 3 we discuss in details the important case of the weak Schwarzschild field
with a spherically symmetric distribution of plasma, and derive the formula for the deflection in this
situation. Our approach allows us consider two effects simultaneously: the difference of the gravitational
deflection in plasma from the vacuum case and the non-relativistic effect (refraction) connected with the
plasma inhomogeneity. In the paper of \cite{Kulsrud-Loeb} it was shown that in the homogeneous plasma the
photon wave packet moves like a particle with a velocity equal to the group velocity of the wave packet, and
with a mass equal to the plasma frequency. In the section 4 we show, that our result for a homogeneous plasma
follows also from this approach. In the section 5 we discuss the observational appearances for a
Schwarzschild point-mass lens, surrounded by a uniform plasma (effect of gravitational radiospectrometer). In
particular, we obtain formulae for the magnification factors in this case. We also estimate a possibility of
observation of this effect by the planned VLBI space project Radioastron. In the section 6 we consider the
models with the nonuniform plasma distribution. For different gravitational lens models we compare the
corrections to the vacuum lensing due to the gravity effect in plasma and due to the plasma inhomogeneity.

After the publication of our previous work \citep{BK&Ts2009} two papers concerning the effect of
amplification of gravitational deflection in dispersive medium were published \citep{medium1, medium2}. These
papers have nothing to do with astrophysics, authors calculate the vertical deflection of a light ray in a
medium with strong frequency-dependent dispersion in a uniform gravitational field, and estimate such effect
for the laboratory experiments with the Earth gravity. In the paper of \citet{medium2} the Synge method of
calculation is used.

\section{Deflection angle in the inhomogeneous plasma in presence of gravity}

Let us consider a static space-time with a metric

\begin{equation}
ds^2 = g_{ik} \, dx^i dx^k = g_{\alpha \beta}  \, dx^\alpha dx^\beta
+ g_{00} \left(dx^0\right)^2 ,
\end{equation}
\[
i,k = 0,1,2,3, \; \; \alpha, \beta = 1,2,3.
\]
Here $g_{ik}$ does not depend on time. Let us assume that the gravitational field is weak, so we can write

\begin{equation}\label{metr}
g_{ik} = \eta_{ik} + h_{ik}, \; \; h_{ik} \ll 1, \; \; h_{ik}
\rightarrow 0 \; \; \mbox{under} \; \; x^\alpha \rightarrow \infty
\, .
\end{equation}
Here $\eta_{ik}$ is the flat space metric $(-1,1,1,1)$, and $h_{ik}$ is a small perturbation. Note
\citep{LL2}, that
\begin{equation}\label{metr1}
g^{ik} = \eta^{ik} - h^{ik}, \quad \eta^{ik} =\eta_{ik},\quad  h^{ik}=h_{ik}.
\end{equation}
Let us consider, in this gravitational field, a static inhomogeneous plasma with a refraction index $n$,
which depends on the space location $x^\alpha$, and the photon frequency $\omega(x^\alpha)$:

\begin{equation} \label{plasma-n}
n^2 = 1 - \frac{\omega_e^2}{[\omega(x^\alpha)]^2}, \quad \omega_e^2 = \frac{4 \pi e^2 N(x^\alpha)}{m} = K_e
N(x^\alpha).
\end{equation}
Here $\omega(x^\alpha)$ is the frequency of the photon, which depends on the space coordinates $x^1$, $x^2$,
$x^3$ due to the presence of the gravitational field (gravitational red shift). We denote $\omega(\infty)
\equiv \omega$, $e$ is the charge of the electron, $m$ is the electron mass, $\omega_e$ is the electron
plasma frequency, $N(x^\alpha)$ is the electron concentration in inhomogeneous plasma, and we do not assume
that $N(\infty) = 0$.

The optics in a curved space-time, in a medium, was developed by \cite{Synge}. It was shown that, for the
static case, the connection between the phase velocity $u$, and a 4-vector of the photon momentum $p^i$,
using of the refraction index of medium $n$, $n=c/u$, is written as

\begin{equation} \label{eq-medium}
\frac{c^2}{u^2}=n^2 = 1 + \frac{p_i p^i}{\left(p^0 \sqrt{-g_{00}}\right)^2} \, .
\end{equation}
Here $c$ is the light velocity in a vacuum. The refraction index
$n$, defined for a plasma in (\ref{plasma-n}), is a function of
$x^\alpha$ and $\omega(x^\alpha)$. In the vacuum $n=1$, and we can
obtain from (\ref{eq-medium}) the usual relation for the square of
the photon 4-vector: $p_i p^i = 0$. In the medium, the square of the
photon 4-vector is not equal to zero. For the medium in a flat
space-time we have
\[
g_{00}=-1, \,\,\, g_{\alpha\alpha}=1,\,\,\, p^0=-p_0, \,\,\,p^\alpha=p_\alpha,
\]
\begin{equation}
n^2 = 1 + \frac{-(p^0)^2 + (p^\alpha)^2}{(p^0)^2},
\end{equation}
and obtain the usual relation between the space and time components of the 4-vector of the photon \citep{LL8,
Zhelezn, Ginzb},

\begin{equation}
(p^\alpha)^2 = n^2 (p^0)^2.
\end{equation}
For a static medium in a static gravitational field,  we have (\citealt{Synge}, see also \citealt{Myoller}):

\begin{equation} \label{Synge}
p_0 \sqrt{-g^{00}} = - p^0 \sqrt{-g_{00}} = - \frac{1}{c} \, \hbar \omega(x^\alpha) \, ,
\end{equation}
where $\hbar$ is the Planck constant. A zero component of  the 4-momentum is the energy divided by $c$ (see
\citealt{LL2}), so in a flat space-time we have:

\begin{equation}
p_0 = - p^0  = - \frac{1}{c} \, \hbar \omega \, ,
\end{equation}
where $\omega \equiv \omega(\infty)$. The components of the 4-vector $p^i$, during an arbitrary motion in a
non-homogeneous medium, in a flat space are written as
\[
p^i = (p^0, p^\alpha) = \left( \frac{\hbar  \omega}{c}, \frac{n
\hbar \omega}{c} \, e^\alpha \right) ,
\]
\begin{equation}
p_i = (p_0, p_\alpha) = \left( - \frac{\hbar \omega}{c},  \frac{n
\hbar \omega}{c} \, e_\alpha \right) , \; \; n^2 = 1 -
\frac{\omega_e^2}{\omega^2},
 \label{eq11}
\end{equation}
where $e^\alpha = p^\alpha / p$ \, and $e_\alpha = p_\alpha/p$ \, ($p = \sqrt{p_1^2 + p_2^2 + p_3^2}$) are
the unit 3-vectors in the direction of the 3-vector $p^\alpha$ and $p_\alpha$ correspondingly. In the flat
space we have $e^\alpha = e_\alpha$. We see that for the photons moving in the plasma  $ p_i p^i = -
m_{eff}^2 c^2$, with $m_{eff}=\frac{\hbar\omega}{c^2}\sqrt{1-n^2}$. Using (\ref{eq-medium}) and
(\ref{Synge}), we see that this relation is valid also for any static gravitational field, with
$m_{eff}=\frac{\hbar\omega(x^\alpha)}{c^2}\sqrt{1-n^2}$. We have then, using (\ref{plasma-n}), that the
effective photon mass $m_{eff}$ and effective velocity $v_{eff}$ in a plasma are written as
(\cite{Kulsrud-Loeb})
\[
m_{eff}=\frac{\hbar\omega(x^\alpha)}{c^2}\sqrt{1-n^2}
=\frac{\hbar\omega_e}{c^2},
\]
\begin{equation}
v_{eff}^2=\frac{p^2c^2}{p^2+m_{eff}^2 c^2}
=\left(1-\frac{\omega_e^2}{\omega^2(x^\alpha)}\right)c^2=n^2c^2.
\label{ph-mass}
\end{equation}
Thus the effective photon velocity equals to the group velocity of the photon in plasma
\begin{equation}
{v_{gr}} = \left(\frac{\partial(n \omega)}{\partial \omega}\right)^{-1} c = n \, c \, .
\end{equation}
The energy of the photon in plasma $E_{eff}$, and relation between the energy and the effective momentum
$p_{eff}$ are also the same, as for the massive particle \citep{MTW}
\begin{equation}
 E_{eff} = \hbar \,\omega(x^\alpha)=\sqrt{-g_{00}} \, c\,p^0,
 \label{mass}
\end{equation}
\[
p_{eff}=\sqrt{\frac{E_{eff}}{c^2}-m_{eff}^2
c^2}=\frac{n\hbar\omega(x^\alpha)}{c}.
\]
The trajectories of photons, in presence of the gravitational field, may be obtained from the variational
principle \citep{Synge}

\begin{equation} \label{var-princ}
\delta \left(\int p_i \, dx^i\right) = 0 ,
\end{equation}
with the restriction  (\ref{eq-medium}), which may be written in the form

\begin{equation} \label{add-cond}
W(x^i,p_i) = \frac{1}{2} \left[ g^{ij} p_i p_j - (n^2-1)  \left(p_0
\sqrt{-g^{00}}\right)^2 \right] = 0.
\end{equation}
Here we define the scalar function $W(x^i,p_i)$ of $x^i$ and  $p_i$.
The variational principle (\ref{var-princ}), with the restriction
$W(x^i,p_i)=0$, leads to the following system of differential
equations \citep{Synge}:

\begin{equation}
\label{D-Eq} \frac{dx^i}{d \lambda} = \frac{\partial W}{\partial
p_i}  \, , \; \; \frac{dp_i}{d \lambda} = - \frac{\partial
W}{\partial x^i} \, ,
\end{equation}
with the parameter $\lambda$  changing along the light trajectory.
In the case of a plasma with the refraction  index (\ref{plasma-n}),
the restriction (\ref{add-cond}) can be reduced, with using of
(\ref{plasma-n}) and (\ref{Synge}), to the form

\begin{equation}
W(x^i,p_i) = \frac{1}{2} \left[ g^{ij} p_i p_j +  \frac{\omega_e^2
\hbar^2}{c^2} \right] = 0.
 \label{eq16}
\end{equation}
From (\ref{D-Eq}) we obtain the system of equations  for the space
components $x^\alpha$, $p_\alpha$:

\begin{equation}  \label{syst-gen}
\frac{dx^\alpha}{d \lambda} = g^{\alpha \beta} p_\beta,  \quad
\frac{dp_\alpha}{d \lambda} = - \frac{1}{2} \, g^{i j}_{, \alpha}
p_i p_j - \frac{1}{2} \frac{\hbar^2}{c^2} \left( \omega_e^2
\right)_{, \alpha} ,
\end{equation}
or
\[
\frac{dx^\alpha}{d \lambda} = g^{\alpha \beta} p_\beta, \quad
\]
\begin{equation} \label{Syst}
\frac{dp_\alpha}{d \lambda} = - \frac{1}{2} \, g^{\beta \gamma}_{,
\alpha}\, p_\beta \,p_\gamma - \frac{1}{2} g^{00}_{, \alpha} \,p_0^2
- \frac{1}{2} \frac{\hbar^2}{c^2} \frac{4 \pi e^2}{m} N_{, \alpha}.
\end{equation}
It follows from (\ref{D-Eq}) that in the static field the component $p_0$ is constant along the trajectory.
Let us consider a photon moving along $z$-axis in a curved space-time, in an inhomogeneous plasma. Due to the
curved space-time metric, and plasma inhomogeneity, the photon will move along the curved trajectory. We use
the approximation, in which deviations of the photon trajectory from the straight line are small. Therefore
for the null approximation we use the 4-vector in a flat space

\begin{equation} \label{null-p}
p^i = \left( \frac{\hbar \omega}{c}, 0, 0,  \frac{n \hbar \omega}{c}
\right)  \, , \quad \, p_i = \left( - \frac{\hbar \omega}{c}, 0, 0,
\frac{n \hbar \omega}{c} \right) .
\end{equation}
The unit 3-vector in the direction of the photon momentum is written in the null approximation as:  $e^\alpha
=e_\alpha= (0,0,1)$. We integrate the equations (\ref{Syst}), calculating the right-hand side, by using the
null approximation in the trajectory of the photon, with $p^i$ from (\ref{null-p}). We obtain than from the
first equation in (\ref{Syst})

\begin{equation}
\frac{dz}{d \lambda} = \frac{n \hbar \omega}{c}, \quad d \lambda  =
c \, \frac{dz}{n \hbar \omega} \, .
\end{equation}
 The deflection angle is determined by a change of 3-vector $e_\alpha$,
 so let us express the second equation in
(\ref{Syst}) through the $e_\alpha$. We obtain successively

\begin{equation}
\frac{n \hbar \omega}{c^2} \, \frac{d(n \hbar \omega  \,
e_\alpha)}{dz} = - \frac{1}{2} \, g^{\beta \gamma}_{, \alpha}
p_\beta p_\gamma - \frac{1}{2} g^{00}_{, \alpha} p_0^2 - \frac{1}{2}
\frac{\hbar^2}{c^2} K_e N_{, \, \alpha},
\end{equation}

\begin{equation}
\frac{d(n \, e_\alpha)}{dz} = e_\alpha \,  \frac{dn}{dz} + n \,
\frac{d e_\alpha}{dz} ,
\end{equation}
\[
n \, \frac{d e_\alpha}{dz} = - e_\alpha \, \frac{dn}{dz} -
\frac{1}{2}  \frac{c^2}{n \hbar^2 \omega^2} \times
\]
\begin{equation}
\times \left( g^{\beta \gamma}_{, \alpha} p_\beta p_\gamma  +
g^{00}_{, \alpha} p_0^2 + \frac{\hbar^2}{c^2} K_e N_{, \, \alpha}
\right).
 \label{ealpha}
\end{equation}
In the right hand side of this equation we use the components from
(\ref{null-p}). We are interested in the components of 3-vector
$e_\alpha$, $e_3=1$, which are orthogonal to the initial direction
of the propagation, at $\alpha=1,2$. In the right hand side of
(\ref{ealpha}) we use the null approximation with $e_\alpha = 0$.
Using (\ref{metr1}), we obtain:
\[
\frac{d e_\alpha}{dz} = \frac{1}{2} \frac{c^2}{n^2 \hbar^2 \omega^2} \times
\]
\begin{equation}
\times \left( h_{33 , \alpha} \frac{n^2 \hbar^2 \omega^2}{c^2}
+h_{00, \alpha} \frac{\hbar^2 \omega^2}{c^2} - \frac{\hbar^2}{c^2}
K_e N_{, \, \alpha} \right) ,
\end{equation}

\begin{equation}
\frac{d e_\alpha}{dz} = \frac{1}{2} \left( h_{33 , \alpha}
+ \frac{1}{n^2} \, h_{00, \alpha}  - \frac{1}{n^2
\omega^2} K_e  N_{, \, \alpha} \right) .
\end{equation}
And for the deflection angle $\hat{\alpha}_\alpha = e_\alpha
(+\infty) - e_\alpha (-\infty)$ we obtain:

\[
\hat{\alpha}_\alpha = \frac{1}{2} \int \limits_{-\infty}^{\infty}
\left( h_{33 , \alpha}  + \frac{h_{00, \alpha}}{1 -
(\omega_e^2/\omega^2)} - \frac{K_e \, N_{, \, \alpha}}{\omega^2 -
\omega_e^2} \right) dz,
\]
\begin{equation}
\alpha = 1,2 .
\end{equation}
For the deflection angle of the photon in the vacuum, $N=0$, we have

\begin{equation}
\hat{\alpha}_\alpha = \frac{1}{2} \int \limits_{-\infty}^{\infty}
\left( h_{33 , \alpha}  +  h_{00, \alpha} \right) dz, \; \; \alpha =
1,2 .
\end{equation}
For the axially symmetric problem, it is convenient to introduce the
impact parameter $b$ relative to the point mass, which remains
constant in the null approximation for the photon moving along the
axis $z$. The plasma has a spherically-symmetric distribution around
the point mass, with the concentration $N = N(r)$. In the axially
symmetric situation the position of the photon is characterized by
$b$ and $z$, and the absolute value of the radius-vector is $r =
\sqrt{x_1^2+x_2^2+z^2}= \sqrt{b^2+z^2}$. We have the following
expression for the deflection angle in the plane perpendicular to
direction of the unperturbed photon  trajectory:
\[
\hat{\alpha}_b = \frac{1}{2} \int  \limits_{-\infty}^{\infty}
\frac{b}{r} \times
\]
\begin{equation} \label{angle}
\times \left( \frac{d h_{33}}{d r}  + \frac{1}{1 -
(\omega_e^2/\omega^2)} \frac{d h_{00}}{d r} - \frac{K_e}{\omega^2 -
\omega_e^2} \frac{d N(r)}{d r} \right) dz .
\end{equation}
Note that $\hat{\alpha}_b < 0$ corresponds to bending of the light trajectory towards the gravitation center,
and $\hat{\alpha}_b > 0$ corresponds to the opposite deflection.

In our previous paper \citep{BK&Ts2009} we have considered a weakly inhomogeneous plasma with  $N(x^\alpha) =
N_0 + N_1(x^\alpha), \; \; N_0 = {\rm const}, \; \; N_1\ll N_0$.  Here we assume that the deflection angle is
small, but we do not assume that $N_1$ is much smaller than $N_0$, so the representation of $N(x^\alpha)$ as
a sum is not required.

\section{Lensing in the Schwarzschild metric}

Let us calculate the deflection angle for a photon moving in the inhomogeneous plasma, in the Schwarzschild
metric of the point mass $M$, with

\begin{equation}
ds^2 = - c^2(1-R_S/r) \, dt^2 + \frac{dr^2}{1-R_S/r}  + r^2(d
\theta^2 + \sin^2 \theta \, d\varphi^2).
\end{equation}
In the weak field approximation this metric is written as \citep{LL2}

\begin{equation}
ds^2 = ds_0^2 + \frac{R_S}{r} (c^2 dt^2 + dr^2),
\end{equation}
where $ds_0^2$ is a flat part of the metric $ds_0^2 =-c^2
dt^2+dr^2+r^2(d \theta^2 + \sin^2 \theta \, d\varphi^2)$. The
components  $h_{ik}$ are written in the Cartesian frame as
\citep{LL2}

\begin{equation}
h_{00} = \frac{R_S}{r}, \quad h_{\alpha \beta} = \frac{R_S}{r}
s_\alpha s_\beta, \quad h_{33} = \frac{R_S}{r} \cos^2 \theta .
\end{equation}
Here $s_\alpha$ is a unit vector in the direction of the radius-vector  $r_\alpha = (x_1,x_2,x_3)$, the
components of which are equal to directional cosines, the angle $\theta$ is the polar angle between 3-vector
$r^\alpha$=$r_\alpha$, and $z$-axis, and $s_3=\cos \theta = z/r = z/\sqrt{b^2+z^2}$. Using formula
(\ref{angle}) we obtain:
\[
\hat{\alpha}_b = - \frac{R_S}{b} \, - \, \frac{1}{2} \int
\limits_{-\infty}^{\infty} \left( \frac{1}{1 -
(\omega_e^2/\omega^2)} \, \frac{R_S \, b}{r^3} \, + \right.
\]
\begin{equation}
\label{angle-inhom} + \left.  \frac{K_e}{\omega^2 -
\omega_e^2}\frac{b}{r}  \frac{d N(r)}{d r} \right) dz .
\end{equation}

To demonstrate the physical meaning of different terms in
(\ref{angle-inhom}),  we write this expression under condition $1-n
= \omega_e^2/\omega^2 \ll 1$. Carrying out the expansion of terms
with the plasma frequency, we obtain:
\[
\hat{\alpha}_b = - \frac{2 R_S}{b} \, - \, \frac{1}{2} \frac{R_S \, b}{\omega^2}
\int\limits_{-\infty}^{\infty} \, \frac{\omega_e^2}{r^3} \, dz \, -
\]
\begin{equation} \label{angle-expansion}
- \frac{1}{2} \, \frac{K_e b}{\omega^2} \int
\limits_{-\infty}^{\infty}    \frac{1}{r} \frac{d N(r)}{d r} \, dz -
\frac{1}{2} \frac{K_e b}{\omega^4} \int \limits_{-\infty}^{\infty}
\frac{\omega_e^2}{r} \, \frac{d N(r)}{d r} \, dz  .
\end{equation}
The first term is a vacuum gravitational deflection. The second term
is  an additive correction to the gravitational deflection, due to
the presence of the plasma. This term is present in the deflection
angle both in the inhomogeneous and in the homogeneous plasma, and
depends on the photon frequency. The third term is a
non-relativistic deflection due to the plasma inhomogeneity (the
refraction). This term depends on the frequency, but it is absent if
the plasma is homogeneous. The forth term is a small additive
correction to the third term.
 If we use the approximation $1-n =
\omega_e^2/\omega^2 \ll 1$, and neglect small second and the forth terms,  we obtain a separate input of the
two effects: the vacuum gravitational deflection, and the refraction deflection in the inhomogeneous plasma.
Calculation of the refraction deflection for a power-law concentration was given in our previous work
\citep{BK&Ts2009}, see also \citet{B-Minakov}, \citet{Muhl1970}, \citet{Thompson}, \citet{zadachnik}. Note
that the refraction in the inhomogeneous plasma with $N(r) = N_0 (R_0/r)^h$, where $N_0$ = const, $R_0$ =
const, $h$ = const $\neq 0$, leads to the refraction deflection angle $\alpha_r$, which is opposite to the
gravitational deflection \citep{B-Minakov, BK&Ts2009}. For $\omega \gg \omega_e$ we have:
\[
\alpha_r = \frac{1}{\omega^2} \frac{4 \pi e^2}{m} N_0
\left(\frac{R_0}{b}\right)^h \frac{\sqrt{\pi} \,
\Gamma\left(\frac{h}{2} +
\frac{1}{2}\right)}{\Gamma\left(\frac{h}{2}\right)} \, ,
\]
\begin{equation}
\Gamma(x) = \int \limits_0^\infty t^{x-1} e^{-t} dt .
\end{equation}
For the arbitrary $n$ one needs to use the expression (\ref{angle-inhom}) which is valid in a general case.
The main approximation used here, is the smallness of the deflection angle, what can be satisfied even if the
concentration $N$ changes significantly, or if the refraction index $n$ is not close to the unity. The most
interesting result following from our calculation is that even in the case of the homogeneous plasma the
photon deflection angle differs from the vacuum case, and depends on the plasma and photon frequency. Indeed,
for $\omega_e $ = const we obtain from (\ref{angle-inhom})

\begin{equation}
\hat{\alpha}_b = - \frac{R_S}{b} \left( 1 + \frac{1}{1 -
(\omega_e^2 / \omega^2)} \right).
\end{equation}
This formula is valid only for $\omega > \omega_e$, because the waves with $\omega< \omega_e$ do not
propagate in the plasma \citep{Ginzb}. Here $\hat{\alpha}_b  < 0$. This means that the light ray is bent in
the direction of the gravitation center, as it occurs in the vacuum. Thus the presence of plasma
\textit{increases} the gravitational deflection angle. This formula is valid under the condition of smallness
of $\hat{\alpha}_b$, but this condition allows the second term in brackets to be much larger than the first
one. So, the gravitational deflection in plasma can be significantly larger than in the vacuum. This effect
has a general relativistic nature, in combination with the dispersive properties of plasma. Such effect may
happen only for radio frequency photons, because optical frequencies are much higher than the plasma
frequency $\omega_e$, so the effect would be negligible.

In the literature on gravitational lensing theory the deflection angle is determined usually as a positive
one (\ref{Einst-angle}), and is defined as the difference between the initial and the final ray directions
$\hat{\bmath{\alpha}} = \bmath{e}_{in} - \bmath{e}_{out}$, where $\bmath{e}$ is the unit tangent vector of a
ray \citep{GL}. Therefore, if we use this definition, we will have the expression with the opposite sign:
\begin{equation} \label{main-res}
\hat{\alpha} = \frac{R_S}{b} \left( 1 + \frac{1}{1 - (\omega_e^2  /
\omega^2)} \right),
\end{equation}
which turns into the deflection angle for vacuum $2R_S/b$, when
$\omega \rightarrow \infty$.

\section{Analogy between a photon in plasma,
             and a massive particle}

Using the analogy between the motion of a photon in the homogeneous plasma, and a massive particle in the
gravitational field (\ref{ph-mass}), (\ref{mass}) we may easily find the deflection angle of the photon in
plasma, produced by a point mass, in a weak field approximation. A test massive particle passing with the
velocity $v$ near a spherical body with a mass $M$, having the impact parameter $b \gg R_S$, deflects to the
angle $\alpha_m$ defined as \citep{MTW, zadachnik}:

\begin{equation}
\alpha_m = \frac{R_S}{b} \left( 1 + \frac{1}{\beta^2} \right),  \;
\;  \beta = \frac{v}{c} .
\end{equation}
If we use in this formula $v=v_{gr} = c \, n = ñ [1-(\omega_e^2/\omega^2)]^{1/2}$  we obtain the same formula
(\ref{main-res}) for the photon deflection in the homogeneous plasma. Note that this analogy is valid for any
static gravitational field, not only for a weak field.

\section{Observational effects. The project Radioastron}

We see from (\ref{main-res}), that photons of smaller frequency, and larger wavelength, are deflected by a
larger angle by the gravitating center. The effect of difference in the gravitational deflection angles is
significant for longer wavelengths, when $\omega$ is approaching $\omega_e$. That is possible only for the
radio waves. Therefore, the gravitational lens in plasma acts as a radiospectrometer \citep{BK&Ts2009}.

The observational effect of the frequency dependence may be represented on the example with off-line lensing
by the Schwarzschild point-mass lens, see fig.1, in the paper of \citep{BK&Ts2009}. This lens gives two
images of the source, on opposite sides of the lens. Angular positions and forms of the images depend on the
Schwarzschild radius of the lens, and on relative positions of the source, the lens, and the observer. The
dependence of the deflection angle on the frequency in plasma leads to smearing of the images, and different
parts of extended images have different spectra. In the case when the source, the lens and the observer are
on the same straight line, the lensing image in vacuum is a thin circle \citep{BK&TsStrong}. In plasma this
circle should have a finite thickness, with the radio spectra depending on the radius. The inner parts of the
circle consist of a more energetic photons than the its outer parts, see fig.2, in the paper of
\citep{BK&Ts2009}. In reality the gravitational lens is not a point mass, it has a complicated structure, and
position of the images differs from that of the point-mass lens. We should also notice that the source must
be radio loud.

The standard model of gravitational lensing is based on the Einstein deflection angle (\ref{Einst-angle}),
which should be replaced, in presence of plasma by our formula (\ref{main-res}).
The angular half separation due to gravitational lensing, between the images of the source in vacuum
\citep{GL} is of the order of

\begin{equation} \label{vac-separ}
\theta_0 = \sqrt{2 R_S \frac{D_{ds}}{D_d D_s}} \, ,
\end{equation}
where $D_d$ is the distance between the observer and the lens, $D_s$ is the distance between the observer and
the source, $D_{ds}$ is the distance between the lens and the source. In the case of a perfect alignment of
the source, the lens and of the observer, the image of the source is called Einstein ring, its radius has an
angular size $\theta_0$. The observed angular separation of quasar images is usually around 1 arcsec for
lensing by a galaxy. Lensing in presence of a homogeneous plasma (\ref{main-res}) leads to an angular half
separation between images as
\[
\theta_0^{pl} = \sqrt{\left( 1 + \frac{1} {1 - (\omega_e^2 /
\omega^2)} \right) R_S \frac{D_{ds}}{D_d D_s}} =
\]
\begin{equation}
= \theta_0 \,\sqrt{ \frac{1}{2} \left( 1 + \frac{1}{1 -  (\omega_e^2
/ \omega^2)} \right)}  \,  \, ,
\end{equation}
which may be called, as plasma Einstein ring. For $\omega_e^2 /
\omega^2 \ll 1$ we obtain

\begin{equation}
\theta_0^{pl} =   \left( 1 + \frac{1}{4} \frac{\omega_e^2}
{\omega^2} \right)  \, \theta_0 \, .
\end{equation}
 The difference between
angular separations of  images in vacuum and in plasma
$\Delta\theta_0$, produced by the same lens configuration, is equal
to

\begin{equation} \label{difference}
\frac{\Delta \theta_0}{\theta_0} = \frac{\theta_0^{pl} -
\theta_0}{\theta_0} = \frac{1}{4} \frac{\omega_e^2}{\omega^2} \simeq
2.0 \cdot 10^7 \, \frac{N_e}{\nu^2} ,
\end{equation}
where $\nu$ is the photon frequency in Hz, $\omega = 2 \pi \nu$. The formula (\ref{difference}) gives the
difference between the deviation angle of the radio wave with a frequency $\nu$, and the optical image, which
may be described by the vacuum formula (\ref{vac-separ}).

Let us estimate the possibility of observation of this effect by the planned project RadioAstron (see the
RadioAstron web page at http://www.asc.rssi.ru/radioastron/index.html). The Radioastron is the VLBI space
project led by the Astro Space Center of Lebedev Physical Institute in Moscow. The payload is the Space Radio
Telescope, based on the spacecraft Spektr-R. For the lowest frequency of the Radioastron, $\nu = 327 \cdot
10^6$ Hz, the angular difference between the vacuum and the plasma images is about $10^{-5}$ arcsec, when the
plasma density on the photon trajectory, in the vicinity of the gravitational lens is of the order of $N_e
\sim 5\cdot 10^4$ cm$^{-3}$. this angular resolution is supposed to be reached in the project Radioastron.

The magnification of the image increases with increasing of the
deflection angle,  therefore different images may have different
spectra in the radio band, when the light propagates in regions with
different plasma density. The magnification is determined by the
deflection law \citep{GL}. Let us demonstrate it on the example of
the point-mass lensing. The magnification factor of the primary
image $\mu_+$, located at the same side as the source relative to
the lens, and of the secondary image $\mu_-$, located at the
opposite side, depend on the angular position of the source.
Corresponding formulae can be found in \cite{GL}
\begin{equation} \label{eqmu+}
\mu_+ = \frac{1}{4} \left[ \frac{y}{\sqrt{y^2+4}} +
\frac{\sqrt{y^2+4}}{y} + 2  \right],
\end{equation}
\begin{equation} \label{eqmu-}
\mu_- = \frac{1}{4} \left[ \frac{y}{\sqrt{y^2+4}} + \frac{\sqrt{y^2+4}}{y} - 2  \right].
\end{equation}
Here $y = \beta / \theta_0$, where $\beta$ is the angular position
of the source relative to the line passing through the observer and
the lens. The total magnification of the source is equal to
\begin{equation}
\mu_{tot} = \mu_+ + \mu_- = \frac{y^2 + 2}{y \sqrt{y^2+4}}.
\end{equation}
Consideration of the total magnification factor is important for the microlensing events when the separated
images are not resolved, and the only observable effect is changing of the flux from the source due to
lensing.

In the case of lensing in plasma we can rewrite these formulae, using $\tilde{y} = \beta/\theta_0^{pl}$
instead $y$:
\begin{equation} \label{mu-pl+}
\mu_+^{pl} = \frac{1}{4} \left[ \frac{\tilde{y}}{\sqrt{\tilde{y}^2+4}} +
\frac{\sqrt{\tilde{y}^2+4}}{\tilde{y}} + 2 \right],
\end{equation}
\begin{equation} \label{mu-pl-}
\mu_-^{pl} = \frac{1}{4} \left[ \frac{\tilde{y}}{\sqrt{\tilde{y}^2+4}} +
\frac{\sqrt{\tilde{y}^2+4}}{\tilde{y}} - 2  \right],
\end{equation}
\begin{equation} \label{mu-pl-tot}
\mu_{tot}^{pl} = \mu_+^{pl} + \mu_-^{pl} = \frac{\tilde{y}^2 +
2}{\tilde{y} \sqrt{\tilde{y}^2+4}},
\end{equation}
\[
\mbox{where} \; \; \tilde{y} = \frac{\beta}{\theta_0^{pl}} =
\frac{\beta}{\theta_0} \left[\frac{1}{2} \left( 1 + \frac{1}{1 -
(\omega_e^2 / \omega^2)} \right)\right]^{-1/2} =
\]
\begin{equation} \label{mu-pl-tot-y}
= y \left[\frac{1}{2} \left( 1 + \frac{1}{1 - (\omega_e^2 /
\omega^2)} \right)\right]^{-1/2} .
\end{equation}
At large $\beta$ the total amplification factor goes to unity, because the influence of the lens on the light
propagation becomes negligible. For the Schwarzschild lens at a small angle $\beta$, the amplification is
inversely proportional to the angle $\beta$, and is proportional to the angular radius of the Einstein ring,
which increases with decreasing frequency, approaching infinity at the plasma frequency. As $\beta$ goes to
zero, the amplification increases, formally unrestrictedly for the point source.

Let us consider the magnification in plasma, when $\omega_e^2 / \omega^2 \ll 1$. Carrying out the expansion
of the expression $(\ref{mu-pl+})$, $(\ref{mu-pl-})$, $(\ref{mu-pl-tot})$, and $(\ref{mu-pl-tot-y})$, we
obtain the additional terms which arise in the case of the plasma, as compared to the case of the vacuum:
\begin{equation}
\mu_+^{pl} = \mu_+ \, + \, \frac{1}{y (y^2+4)^{3/2}} \,
\frac{\omega_e^2}{\omega^2},
\end{equation}
\begin{equation}
\mu_-^{pl} = \mu_- \, + \, \frac{1}{y (y^2+4)^{3/2}} \,
\frac{\omega_e^2}{\omega^2} ,
\end{equation}
\begin{equation}
\mu_{tot}^{pl} = \mu_{tot} \, + \, \frac{2}{y (y^2+4)^{3/2}} \,
\frac{\omega_e^2}{\omega^2} .
\end{equation}
We see that the presence of a homogeneous plasma increases the magnification.

\begin{figure}
\centerline{\hbox{\includegraphics[width=0.48\textwidth]{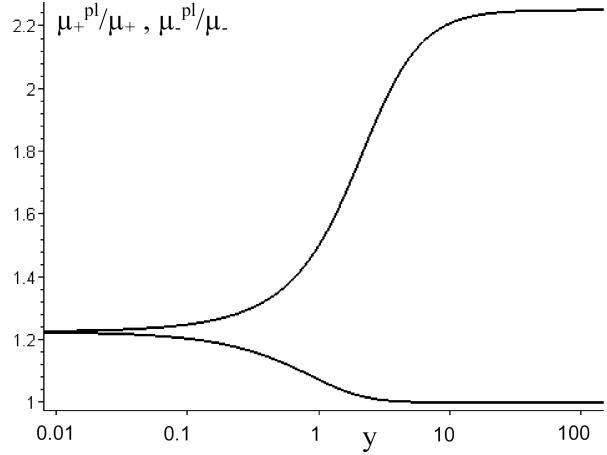}}} \caption{The ratio $\mu_+^{pl}/\mu_+$ of
the magnification of the primary image in presence of plasma to the same value in vacuum (lower curve), and
the ratio $\mu_-^{pl}/\mu_-$ of the magnification of the secondary image in presence of plasma to the same
value in vacuum (upper curve). Curves are plotted for $\omega=\sqrt{2} \, \omega_e$.}
 \label{mmag}
\end{figure}
The ratio of the magnification of the primary image in presence of plasma $\mu_+^{pl}$ to the same value in
vacuum $\mu_+$ is given in Fig.\ref{mmag}, for $\omega=\sqrt{2} \, \omega_e$, according to (\ref{eqmu+}),
(\ref{mu-pl+}). The upper curve in Fig.1 is the ratio of the magnification of the secondary image in presence
of plasma $\mu_-^{pl}$ to the same value in vacuum $\mu_-$, according to (\ref{eqmu-}), (\ref{mu-pl-}). As
$y$ approaches zero, the ratio of the magnifications $\mu_+^{pl}/\mu_+$ and $\mu_-^{pl}/\mu_-$ becomes
constant $V_{+}$, depending on the frequency
\begin{equation}
V_{+} = \frac{\mu_+^{pl}}{\mu_+} = \frac{\mu_-^{pl}}{\mu_-} = \frac{\theta_0^{pl}}{\theta_0}= \sqrt{
\frac{1}{2} \left( 1 + \frac{1}{1 - (\omega_e^2 / \omega^2)} \right)}.
\end{equation}
This value is equal $\sqrt{ \frac{3}{2}}$ at  $\omega=\sqrt{2} \, \omega_e$. At large $y$ the ratio
$\mu_+^{pl}/\mu_+$ is tending to unity, the ratio $\mu_-^{pl}/\mu_-$ is tending to the constant $\tilde{V}_-$
\begin{equation}
\tilde{V}_- \approx \left(\frac{\theta_0^{pl}}{\theta_0}\right)^4 = \left[ \frac{1}{2} \left( 1 + \frac{1}{1
- (\omega_e^2 / \omega^2)} \right) \right]^2.
\end{equation}
This value is equal $9/4$ at  $\omega=\sqrt{2} \, \omega_e$. Note, that both $\mu_-$ and $\mu_-^{pl}$ are
tending to zero at this limit.

The deflection angle in presence of the plasma is larger, than in the vacuum, so the amplification for lower
frequencies is larger, and in this situation the image spectrum differs from the original spectrum of the
source, having more intensive the low-frequency part. The light in two lensing images is propagating through
different media with different plasma density. Therefore, different images of the same source may have
different spectra in the radio band.

In the ideal case, when two images of the same source are formed by rays propagating through the uniform
plasma with different concentrations, the spectra of two images should be different. In the achromatic
lensing the ratio of fluxes should be the same at all frequencies $\mu_+^{opt}=\mu_+^{rad}$,
$\mu_-^{opt}=\mu_-^{rad}$, so that $\mu_+^{opt}/\mu_-^{opt}= \mu_+^{rad}/\mu_-^{rad}$. The presence of plasma
leads to larger amplification at lower frequencies, so if the ratio of fluxes in optics and in lower radio
band of two lensing images is different $\mu_+^{opt}/\mu_-^{opt} \neq \mu_+^{rad}/\mu_-^{rad}$, it may be
related to plasma influence, so that the image with larger relative radio flux propagates through the plasma
with larger density. In Fig.2 two cases are shown for the value $(\mu_+^{pl}/\mu_+)/(\mu_-^{pl}/\mu_-)$, as a
function of $\omega/\omega_e$, for different fixed $y$. In the first one the plasma density is two times
larger for rays forming the main image, and in the second case the ratio of densities is opposite. While the
vacuum value $\mu_+/\mu_-$ does not depend on the frequency, one of these two behaviors is expected for the
ratio of fluxes of two images as a function of the frequency. In the considered ideal case this plot should
give the information about the ratio of plasma densities, and when the image with lower relative radio flux
is formed by vacuum lensing, it would be possible to obtain the absolute value of the plasma density through
which propagate rays from the image with higher relative radio flux.

From Fig. 2 we see that the spectral dependences can have different forms for different $y$: while the dashed
line is less than unity, the solid line can be both larger and less than unity. In Fig.2a in case when the
light rays corresponding to primary image go through the densier plasma, we see that for smaller frequencies
the primary image is more magnificated, relative to the vacuum magnification, due to the plasma presence than
the secondary image, see solid line. In opposite case, when the light rays corresponding to the secondary
image pass through greater density, we see that for the small frequencies the secondary image is more
magnificated, therefore the dashed line is less than unity. For bigger $y$ both curves are less than unity.
In Fig.2b,c we see that with increasing of $y$ the solid line goes to the region, which is lower than unity.
In Fig.3 we plot the same dependence for the case, when the plasma density is ten times larger for rays
forming the main image, and the second curve corresponds to the opposite ratio in densities. In the case,
when the difference between densities is larger (comparing with situation in Fig.2), effects connected with
plasma are greater. Here the solid line remains above unity for greater values of $y$ than for the situation
described in Fig.2. We should note, that for $y \geq 1$ the secondary image is very demagnified \citep{GL,
GL2}, so in a real situation we will have behaviour like on Fig.2a and Fig3a, and ratio of densities can be
uniquely defined. More detail can be found in Appendix 1.

In this section we have discussed in details new observational effects, connected with the gravitational
deflection in a homogeneous plasma. In reality the distribution of plasma around the gravitating objects is
non-homogeneous, and there is a deflection connected with the medium inhomogeneity. For the plasma density
profile decreasing with the distance from center, the refraction deflection has a sign opposite to the
gravitational one. Therefore the refraction and gravitational deflection in the nonuniform plasma partially
cancel each other, and the value and the sign of the resulting angle depend on the particular configuration
of the gravitational lens. In the following section we consider situations of lensing with the non-uniform
plasma distribution.

\begin{figure}
\centerline{\hbox{\includegraphics[width=0.48\textwidth]{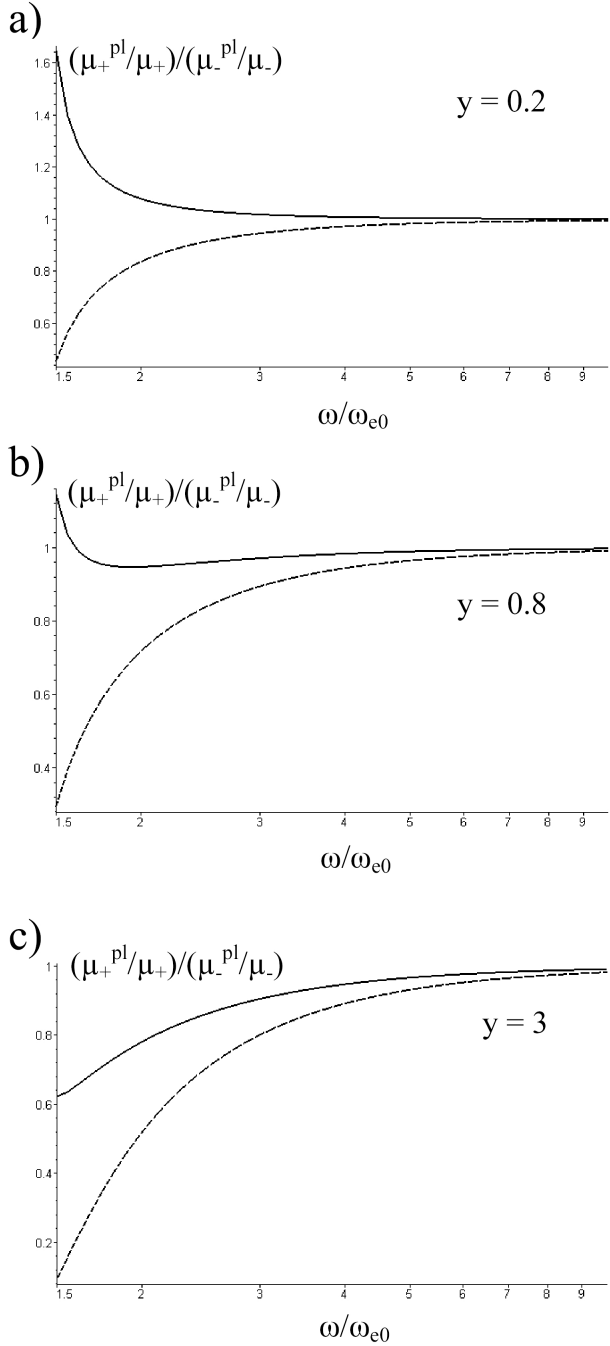}}} \caption{The value
$(\mu_+^{pl}/\mu_+)/(\mu_-^{pl}/\mu_-)$, as a function of $\omega/\omega_e$, for different fixed $y$.
$\omega_{e+}$ is the plasma frequency corresponding to the plasma concentration $N_+$, which rays forming
primary image go through. $\omega_{e-}$ is the plasma frequency corresponding to the plasma concentration
$N_-$, which rays forming primary image go through. The solid curve is plotted for $\omega_{e+} = \omega_{e0}
\sqrt{2}$, $\omega_{e-} = \omega_{e0}$ ($\omega_{e0}$ is a dimensional constant), it corresponds to $N_+/N_-
= 2$. The dashed curve is plotted for $\omega_{e+} = \omega_{e0}$, $\omega_{e-} = \omega_{e0} \sqrt{2}$, it
corresponds to $N_+/N_- = 1/2$.}
\end{figure}

\begin{figure}
\centerline{\hbox{\includegraphics[width=0.48\textwidth]{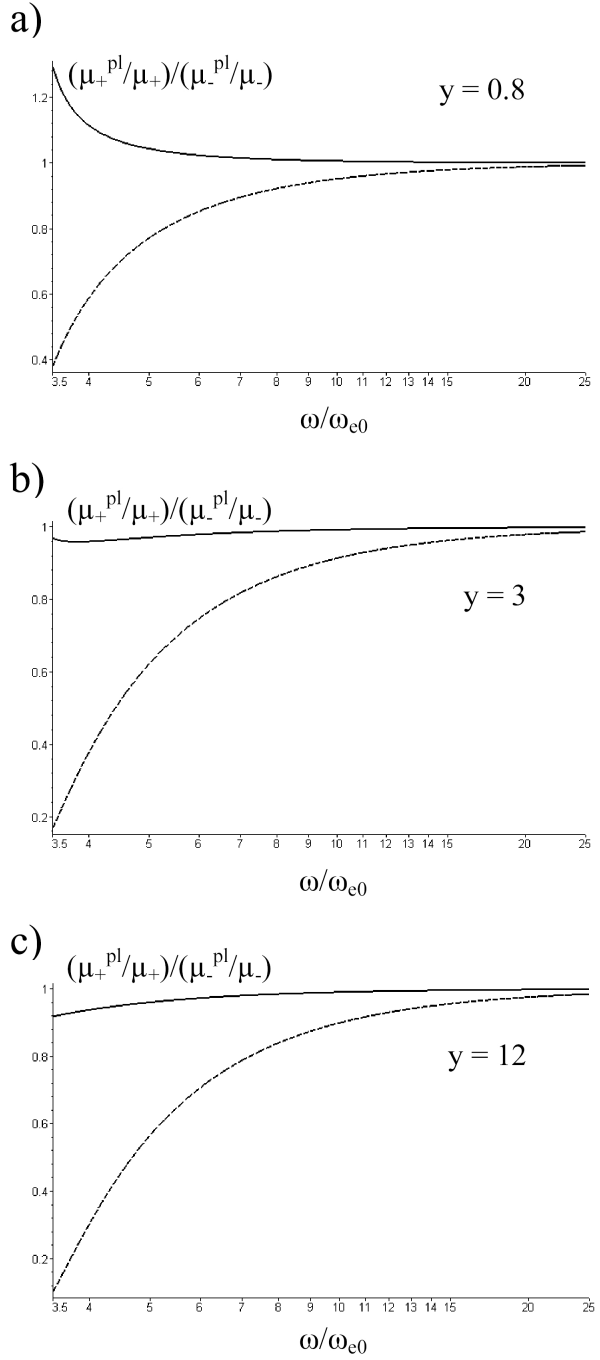}}} \caption{The value
$(\mu_+^{pl}/\mu_+)/(\mu_-^{pl}/\mu_-)$, as a function of $\omega/\omega_e$, for different fixed $y$. The
solid curve is plotted for $\omega_{e+} = \omega_{e0} \sqrt{10}$, $\omega_{e-} = \omega_{e0}$ ($\omega_{e0}$
is a dimensional constant), it corresponds to $N_+/N_- = 10$. The dashed curve is plotted for $\omega_{e+} =
\omega_{e0}$, $\omega_{e-} = \omega_{e0} \sqrt{10}$, it corresponds to $N_+/N_- = 1/10$.}
\end{figure}

\section{Models with nonuniform plasma distribution}

\subsection{Singular isothermal sphere}

Let us consider a simple model of singular isothermal sphere
\citep{BinneyTremaine, Chandrasekhar} which is often used in the
lens modeling of galaxies and clusters of galaxies \citep{GL, GL2,
Bartelmann-Schneider, Wambsganss}. The density distribution is
written as

\begin{equation} \label{SIS-rho}
\rho(r) = \frac{\sigma_v^2}{2 \pi G r^2}.
\end{equation}
Here $\sigma_v$ is a one-dimensional velocity dispersion (for stars
in galaxies or for galaxies in clusters of galaxies). The projected
surface mass density of a singular isothermal sphere to the lens
plane, perpendicular to the light ray, is equal to

\begin{equation}
\Sigma(b) = \frac{\sigma_v^2}{2 G b}.
\end{equation}
The deflection due to a  axisymmetric mass distribution with the impact parameter $b$ is equal to the
Einstein angle for the mass $M(b)$, where $M(b)$ is the projected mass enclosed by the circle of the radius
$b$. In another words it is the mass inside the cylinder with the radius $b$ \citep{Clark, B-Minakov, GL,
GL2}. So we for this case we should rewrite formula (\ref{angle-expansion}) substituting $M(b)$:

\[
\hat{\alpha}_b = \frac{4 G M(b)}{c^2 b} \, + \, \frac{2 G M(b) \,
b}{c^2 \omega^2} \int\limits_{0}^{\infty} \, \frac{\omega_e^2}{r^3}
\, dz \, +
\]
\begin{equation} \label{alphy}
+ \, \frac{K_e b}{\omega^2} \int \limits_{0}^{\infty}  \frac{1}{r}
\frac{d N(r)}{d r} \, dz + \frac{K_e b}{\omega^4} \int
\limits_{0}^{\infty} \frac{\omega_e^2}{r} \, \frac{d N(r)}{d r} \,
dz =
\end{equation}
\[
= \hat{\alpha}_1 + \hat{\alpha}_2 + \hat{\alpha}_3 + \hat{\alpha}_4 .
\]
Here we rewrite the formula in such a way that the gravitational deflection is positive as it is usually
determined in the gravitational lensing theory (compare with (\ref{angle-expansion})). The term
$\hat{\alpha}_1$ is the vacuum gravitational deflection, the term $\hat{\alpha}_2$ is the correction to the
gravitational deflection, due to the presence of the plasma, the term $\hat{\alpha}_3$ is the refraction
deflection due to the inhomogeneity of the plasma, the term $\hat{\alpha}_4$ is a correction to the third
term. We are interested mainly in the effects, described by the terms $\hat{\alpha}_1$, $\hat{\alpha}_2$ and
$\hat{\alpha}_3$.

For the singular isothermal sphere the projected mass $M(b)$ is:
\begin{equation}
M(b) = \int \limits_0^{b} \Sigma(b')  \, 2 \pi b' db' =
\frac{\sigma_v^2}{G} \, \pi b.
\end{equation}
So the vacuum gravitational deflection due to the singular
isothermal sphere is constant:
\begin{equation}
\hat{\alpha}_1 = 4 \pi \frac{\sigma_v^2}{c^2}.
\end{equation}
For the concentration of the plasma we have:
\begin{equation} \label{NcherezRho}
N(r) = \frac{\rho(r)}{\kappa m_p},
\end{equation}
where $m_p$ is the proton mass, and $\kappa$ is a non-dimensional coefficient responsible for the dark matter
contribution, and is approximately equal $\kappa \simeq 6$. We assume here that $\rho(r)$, which is given by
the formula (\ref{SIS-rho}), is the density of all kind of matter, not only plasma particles. Thus, the
plasma frequency is equal to
\begin{equation}
\omega_e^2 = K_e N(r) = \frac{K_e}{\kappa m_p} \, \rho(r).
\end{equation}
The correction to the gravitational deflection due to the presence of the plasma is:
\[
\hat{\alpha}_2 = 2 \pi \frac{b^2 \sigma_v^2}{\omega^2 c^2}
\frac{K_e}{\kappa m_p} \int \limits_0^{\infty} \frac{\rho(r)}{r^3}
dz =
\]

\begin{equation}
= \frac{b^2 \sigma_v^2}{\omega^2 c^2} \frac{K_e}{\kappa m_p}
\frac{\sigma_v^2}{G} \int \limits_0^{\infty}
\frac{dz}{(b^2+z^2)^{5/2}} .
\end{equation}
Integration of such expressions can be performed using \citet{Gr-R},
and the properties of the $\Gamma$-function:

\begin{equation} \label{step-integral}
\int \limits_{0}^{\infty} \frac{dz}{(z^2+b^2)^{h/2+1}}
= \frac{1}{h b^{h+1}} \frac{\sqrt{\pi} \,
\Gamma\left(\frac{h}{2} + \frac{1}{2}\right)}{\Gamma\left(\frac{h}{2}\right)} ,
\end{equation}
\[
\mbox{where} \; \; \Gamma(x) = \int \limits_0^\infty t^{x-1} e^{-t} dt .
\]
Using (\ref{step-integral}), we obtain the deflection angles
$\hat{\alpha}_2$ and $\hat{\alpha}_3$:
\begin{equation}
\hat{\alpha}_2 = \frac{2}{3} \, \frac{\sigma_v^2}{c^2} \,
\frac{K_e}{\kappa m_p} \, \frac{\sigma_v^2}{G
\omega^2 b^2} ,
\end{equation}
\begin{equation}
\hat{\alpha}_3 = - \frac{K_e b}{\omega^2 \kappa m_p}
\frac{\sigma_v^2}{\pi G} \int \limits_0^\infty
\frac{dz}{(z^2+b^2)^2} = - \frac{1}{4} \, \frac{K_e}{\kappa m_p}
\, \frac{\sigma_v^2}{G \omega^2 b^2} .
\end{equation}
For the ratio of the angles $\hat{\alpha}_2$ and $\hat{\alpha}_3$ we have:
\begin{equation}
\left| \frac{\hat{\alpha}_2}{\hat{\alpha}_3} \right|
= \frac{8}{3} \frac{\sigma_v^2}{c^2}.
 \label{ratio}
\end{equation}

\subsection{Non-singular isothermal gas sphere}

Let us consider a gravitational lens model of an isothermal sphere,
in which the singularity at the origin is replaced by a finite core
\citep{Hinshaw-Krauss, Wu}:
\begin{equation} \label{ISCmodel}
\rho(r) = \frac{\sigma_v^2}{2 \pi G (r^2+r_c^2)} =
\frac{\rho_0}{\left( 1+\frac{r^2}{r_c^2} \right)^2}, \quad \rho_0 =
\frac{\sigma_v^2}{2 \pi G r_c^2},
\end{equation}
where $r_c$ is the core radius. The corresponding projected surface
mass density for this model is
\begin{equation}
\Sigma(b) = \frac{\sigma_v^2}{2G \sqrt{b^2+r_c^2}}.
\end{equation}
The total projected mass within $b$ and the vacuum gravitational deflection angle are:
\begin{equation}
M(b) = \pi \frac{\sigma_v^2}{G} \left( \sqrt{b^2+r_c^2} - r_c \right),
\end{equation}
\begin{equation}
\hat{\alpha}_1 = 4 \pi \frac{\sigma_v^2}{c^2} \frac{\sqrt{b^2+r_c^2} - r_c}{b}.
\end{equation}
Analogically to the previous subsection, we reduce the angle $\hat{\alpha}_2$ to the form:

\begin{equation}
\hat{\alpha}_2 = \frac{2 G M(b) \, b}{c^2 \omega^2} \,
\frac{K_e}{\kappa m_p} \, \frac{\sigma_v^2}{2 \pi G} \int
\limits_0^\infty \frac{dz}{r^3 (r^2+r_c^2)} =
\end{equation}
\[
= \frac{2 G M(b) \, b}{c^2 \omega^2} \, \frac{K_e}{\kappa m_p} \,
\frac{\sigma_v^2}{2 \pi G} \int \limits_0^\infty
\frac{dz}{(z^2+b^2)^{3/2} (z^2+b^2+r_c^2)}.
\]
By substitution $z^2/(z^2+b^2)=x^2$ the integral
\[
I_1 = \int \limits_0^\infty \frac{dz}{(z^2+b^2)^{3/2}
(z^2+b^2+r_c^2)}
\]
is reduced to
\begin{equation}
I_1 = \int \limits_0^1 \left( \frac{1}{b^2 r_c^2} - \frac{1}{r_c^2
(b^2+r_c^2 (1-x^2))} \right) dx.
\end{equation}
Integration can be performed with using of \citet{Dwight}:
\begin{equation}
I_1 = \frac{1}{b^2 r_c^2} - \frac{1}{r_c^3 \sqrt{b^2+r_c^2}} \,
\arctanh \left( \frac{r_c}{\sqrt{b^2+r_c^2}} \right) .
\end{equation}
For $\hat{\alpha}_2$ we obtain:
\begin{equation}
\hat{\alpha}_2 = \frac{\sigma_v^4}{c^2 \omega^2 G} \,
\frac{K_e}{\kappa m_p} \, b \left( \sqrt{b^2+r_c^2} - r_c \right)
\times
\end{equation}
\[
\times \left[ \frac{1}{b^2 r_c^2} - \frac{1}{r_c^3 \sqrt{b^2+r_c^2}}
\, \arctanh \left( \frac{r_c}{\sqrt{b^2+r_c^2}} \right) \right] .
\]
Angle $\hat{\alpha}_3$ can be calculated using formula
(\ref{step-integral}):
\begin{equation}
\hat{\alpha}_3 = - \frac{K_e b}{\omega^2} \frac{1}{\kappa m_p} \,
\frac{\sigma_v^2}{2 \pi G} \int \limits_0^\infty
\frac{dz}{(z^2+b^2+r_c^2)^2} =
\end{equation}
\[
= - \frac{1}{4} \frac{\sigma_v^2}{\omega^2 G} \,  \frac{K_e}{\kappa
m_p} \, \frac{b}{(b^2+r_c^2)^{3/2}}.
\]
Let us express angles $\hat{\alpha}_2$ and $\hat{\alpha}_3$ in terms of the central density $\rho_0$, using
(\ref{ISCmodel}). We obtain for $r_c \gg b$:

\begin{equation}
\hat{\alpha}_2 = \frac{2 \pi^2 G \rho_0^2}{c^2 \omega^2}
\frac{K_e}{\kappa m_p} \, b r_c, \quad \hat{\alpha}_3 = - \frac{\pi
\rho_0}{2 \omega^2} \frac{K_e}{\kappa m_p} \, \frac{b}{r_c} \, ;
\end{equation}
and for $b \gg r_c$, similar to the case of the singular isothermal
sphere, we have

\begin{equation}
\hat{\alpha}_2 = \frac{8 \pi^2 G \rho_0^2}{3 c^2 \omega^2}
\frac{K_e}{\kappa m_p} \, \frac{r_c^4}{b^2}, \quad \hat{\alpha}_3 =
- \frac{\pi \rho_0}{2 \omega^2} \frac{K_e}{\kappa m_p} \,
\frac{r_c^2}{b^2} \, .
\end{equation}
Introducing the mass of the uniform core
$M_c=\frac{4\pi}{3}\rho_0r_c^3$, and its gravitational radius
$R_{sc}=\frac{2GM_c}{c^2}$, we obtain the ratio of these angles as

\begin{equation}
\left| \frac{\hat{\alpha}_2}{\hat{\alpha}_3} \right| =
\frac{3R_{sc}}{2r_c}\,\,\,(r_c \gg b),\quad \left|
\frac{\hat{\alpha}_2}{\hat{\alpha}_3} \right| =
\frac{2R_{sc}}{r_c}\,\,\,(r_c \ll b).
 \label{ratio1}
\end{equation}
In the realistic cases  $\left|
\frac{\hat{\alpha}_2}{\hat{\alpha}_3} \right|\ll 1$ in
(\ref{ratio}),(\ref{ratio1}), because spheres have $\sigma_v\ll c$,
and $R_{sc}\ll r_c$. Besides, these relations are needed for the
stability of isothermal spheres, see \cite{BKZ}. Therefore in this
configuration the nonuniform plasma deflection effects are much
stronger, than the gravitational plasma effects, and have an
opposite direction.

\subsection{Plasma sphere around a black hole}

 Let us consider a black hole of mass $M_0$, surrounded by the
electron-proton plasma. We will consider a case, when we can neglect the self-gravitation of the plasma
particles, compared to the gravity of a central black hole. Let us find, in the newtonian approximation, a
density distribution of the plasma in the gravitational field a central point mass $M_0$. The equation of
hydrostatic equilibrium for spherically symmetric mass distribution of a isothermal gas with the equation of
state $P = \rho \Re T$ in the field of the central mass is \citep{BinneyTremaine, Chandrasekhar}

\begin{equation} \label{hydrostM0}
\frac{\Re \, T}{\rho} \frac{d\rho}{dr} = - \frac{G M_0}{r^2}.
\end{equation}
Here $\rho$, $P$ and $T$ are the density, the pressure and the temperature of the plasma, $\Re = k_B/m_p$ is
the gas constant, $k_B$ is the Boltzmann's constant and $m_p$ is the proton mass. To obtain a deflection
angles we need to find the plasma density distribution from the equation (\ref{hydrostM0}) and calculate the
integrals with this density in the formula (\ref{alphy}). But it is interesting that we can find the ratio of
angles $|\hat{\alpha}_2 / \hat{\alpha}_3|$ without solving the equation (\ref{hydrostM0}) and calculation of
the integrals in $\hat{\alpha}_2$ and $\hat{\alpha}_3$. Let us rewrite the equation (\ref{hydrostM0}) to the
form:

\begin{equation}
\frac{K_e b}{\omega^2} \, \Re \, T \frac{1}{r}  \frac{dN(r)}{dr} = -
\frac{K_e b}{\omega^2} \, \frac{G M_0}{r^3} N(r),
\end{equation}
where $N(r)=\rho(r)/m_p$ is the plasma concentration. If we compare it with expressions for $\hat{\alpha}_2$
and $\hat{\alpha}_3$ in (\ref{alphy}) and take into account that $\omega_e^2 = (4 \pi e^2/m) N(r) = K_e
N(r)$, we obtain that

\begin{equation}
\left| \frac{\hat{\alpha}_2}{\hat{\alpha}_3} \right| = \frac{2 \Re T}{c^2}.
\end{equation}
We see that this ratio does not depend on the value of the central mass $M_0$.
 For the boundary condition

\begin{equation}
\rho(r_0) = \rho_0
\end{equation}
we obtain from the equation (\ref{hydrostM0})the density
distribution

\begin{equation}
\rho(r) = \rho_0 \, e^{\frac{GM_0}{\Re \, T} \left( \frac{1}{r} -
\frac{1}{r_0} \right)} = \rho_0 \,
e^{B\left(\frac{1}{r}-\frac{1}{r_0}\right)}, \; \; B =
\frac{GM_0}{\Re \, T}.
\end{equation}
For angles $\hat{\alpha}_2$ and $\hat{\alpha}_3$ we obtain:

\begin{equation}
\hat{\alpha}_2 = \frac{2 G M_0 b}{c^2 \omega^2}  \frac{K_e}{m_p} \,
\rho_0 \, e^{-\frac{B}{r_0}} Int2,
\end{equation}
\begin{equation}
\hat{\alpha}_3 = - \frac{K_e b}{\omega^2 m_p}  \, B \, \rho_0 \,
e^{-\frac{B}{r_0}} \, Int2,
\end{equation}

\begin{equation}
\mbox{where} \; \; Int2 = \int \limits_0^{\infty}
\frac{1}{(b^2+z^2)^{3/2}} \, e^{\frac{B}{\sqrt{b^2+z^2}}} \, dz .
\end{equation}
Calculation of $Int2$ can be found in Appendix 2.

While the temperature of the non-self-gravitating sphere may have arbitrary values, the plasma effects may be
comparable, and even less than the GR plasma effects. It is due to the fact, that with increasing temperature
the plasma density can become arbitrary uniform, with corresponding decreasing of non-uniform plasma effect
for refraction. We have used newtonian non-relativistic description of the gas sphere, but from the arguments
listed above it is clear, that this conclusion remains valid also in the correct relativistic consideration.

\subsection{Plasma in a galaxy clusters}

In a galaxy cluster the electron distribution may be more homogeneous due to large temperature of electrons.
An appropriate approach for this case is to consider a singular isothermal sphere as a model for the
distribution of the gravitating matter, neglecting the mass of plasma, and to find a plasma density
distribution from the solution of the equation of the hydrostatic equilibrium of plasma in the gravitational
field of a singular isothermal sphere. In this approximation the density distribution of the gravitating
matter has a form

\begin{equation}
\rho_{gr}(r) = \frac{\sigma_v^2}{2 \pi G r^2},
\end{equation}
with the vacuum gravitational deflection angle

\begin{equation}
\hat{\alpha}_1 = 4 \pi \frac{\sigma_v^2}{c^2}.
\end{equation}

For the mass inside a sphere with a radius $r$ we have

\begin{equation}
M(r) = \frac{2 \sigma_v^2}{G} \, r ,
\end{equation}
and the equation of a hydrostatic equilibrium of the plasma is:

\begin{equation} \label{hydrostSIS}
\frac{\Re \, T}{\rho} \frac{d\rho}{dr} = - \frac{2 \sigma_v^2}{r}.
\end{equation}
For the boundary condition $\rho(r_0) = \rho_0$ we obtain the plasma
density from the equation (\ref{hydrostSIS}) as

\begin{equation}
\rho(r) = \rho_0 \left( \frac{r}{r_0} \right)^{-s},  \quad
s=\frac{2 \sigma_v^2}{\Re T}.
\end{equation}
The angles $\hat{\alpha}_2$ and $\hat{\alpha}_3$ are calculated using formula (\ref{step-integral}), so we
have

\begin{equation}
\hat{\alpha}_2 = \frac{2 \sigma_v^2 \pi \sqrt{\pi}}{c^2  \omega^2
(s+1)} \frac{K_e}{m_p} \, \frac{\Gamma(\frac{s}{2} +
1)}{\Gamma(\frac{s+1}{2})} \, \rho_0 \, \left( \frac{r_0}{b}
\right)^s ,
\end{equation}
\begin{equation}
\hat{\alpha}_3 = - \frac{K_e b}{\omega^2} \frac{1}{m_p}  \, \rho_0
\, r_0^s \, s \int \limits_0^\infty \frac{dz}{(z^2+b^2)^{s/2 + 1}} =
\end{equation}
\[
= - \frac{K_e \sqrt{\pi}}{\omega^2 m_p} \, \frac{\Gamma(\frac{s}{2} + \frac{1}{2})}{\Gamma(\frac{s}{2})} \,
\rho_0 \, \left( \frac{r_0}{b} \right)^s .
\]
The ratio of these angles is equal to

\begin{equation}
\left| \frac{\hat{\alpha}_2}{\hat{\alpha}_3} \right| =
\frac{\sigma_v^2}{c^2} \, \pi \, \frac{\Gamma(\frac{s}{2} + 1)
\Gamma(\frac{s}{2})}{\Gamma(\frac{s}{2} + \frac{3}{2})
\Gamma(\frac{s}{2} + \frac{1}{2})} .
\end{equation}
Under the condition $s \ll 1$ what corresponds to  $2 \sigma_v^2 \ll
\Re T$, the expressions are simplified to

\begin{equation}
\hat{\alpha}_2 = 2 \pi \frac{\sigma_v^2}{c^2 \omega^2}
\frac{K_e}{m_p} \, \rho_0 \left( \frac{r_0}{b} \right)^s ,
\end{equation}

\begin{equation}
\hat{\alpha}_3 = - \pi \frac{\sigma_v^2}{\Re T \omega^2}
\frac{K_e}{m_p} \, \rho_0 \left( \frac{r_0}{b} \right)^s ,
\end{equation}

\begin{equation}
\left| \frac{\hat{\alpha}_2}{\hat{\alpha}_3} \right|  = \frac{2 \Re
T}{c^2} .
\end{equation}
If relativistic plasma is present in a galaxy cluster, for example in jets from AGNs, the plasma GR effects
may be larger than the effects of the nonuniform plasma. If the distribution of plasma is not spherically
symmetric, there may be distribution of plasma with the density gradient opposite to the direction of the
gravitational force, for example, in the presence of rotation. In this situation the angles $\hat{\alpha}_2$
and $\hat{\alpha}_3$ may be of the same sign.

Let us estimate Thomson optical depth for Thomson scattering of photons in the case of gravitational lensing
in plasma, and depth of the bremsstrahlung absorption. For Thomson optical depth $\tau$ we have:
\begin{equation}
\tau \approx N \sigma_e L,
\end{equation}
where $N$ is the electron concentration, $\sigma_e = 6.65 \cdot 10^{-25}$ cm$^2$ is Thomson scattering cross
section, $L$ is characteristic distance for scattering. So the optical depth will be $\tau < 1$ for $N \leq
0.5 \cdot 10^4 \left(\frac{10^2 \mbox{ pc}}{L}\right)$ cm$^{-3}$.

For optical depth for the absorption of radiation in case $\omega \gg \omega_e$ we have \citep{Ginzb}:
\begin{equation}
\tau = \frac{10^{-2} N^2}{T^{3/2} \nu^2} \left[ 17.7 + \ln \frac{T^{3/2}}{\nu} \right] L .
\end{equation}
So the optical depth will be $\tau < 1$ for
\begin{equation}
N \simeq 0.2 \cdot 10^3 \left( \frac{T}{10^4 \mbox{ K}} \right)^{3/4} \frac{\nu}{10^9 \mbox{ Hz}} \left(
\frac{10^2 \mbox{ pc}}{L} \right)^{1/2} .
\end{equation}

\section{Conclusions}

In an observation of two images of the lensing point source in presence of plasma, the following features may
be observed.

1. Spectra of two images may be different in the long wave side due to different plasma properties along the
trajectories of light rays forming the images.

2. The extended image may have different spectra in different parts
of the image, with a maximum of the spectrum shifting to the long
wave side in the regions with a larger deflection angle.

3. The presence of plasma may influence on the timing effects in
binary relativistic systems, similar to the double pulsar system
J0737-3039A - J0737-3039B, and may induce the spectral dependence of
the properties of the fluctuations of the microwave background
radiation.

4. We have carried out the calculations for models with the nonuniform plasma distribution: singular and
non-singular isothermal sphere; for hot gas inside the gravitational field of a black hole, and of a cluster
of galaxies.

5. For different gravitational lens models we compare the corrections to the vacuum lensing due to the
gravity effect in plasma, and due to the plasma inhomogeneity. We have shown that the gravitational effect
could be detected in the case of a hot gas in the gravitational field of a galaxy cluster.

6. We made estimations of the optical depth due to Thomson scattering and free-free absorption in the process
of during the gravitational lensing in plasma.

\section*{Appendix 1. Investigation of spectral behaviour in Fig.2,3.}
Value $(\mu_+^{pl}/\mu_+)/(\mu_-^{pl}/\mu_-)$ under $\omega \gg \omega_e$ has the following form:

\begin{equation}
(\mu_+^{pl}/\mu_+)/(\mu_-^{pl}/\mu_-) = 1 + \frac{1}{2 \omega^2 (y^2 + 4)} \times
\end{equation}
\[
\times \left[ \omega_{e+}^2 \left(y^2 + 2 - y \sqrt{y^2+4}\right) - \omega_{e-}^2 \left(y^2 + 2 + y
\sqrt{y^2+4}\right) \right].
\]

We see that:

a) if the rays forming the primary image go through greater density than the rays forming the secondary
image, that is $\omega_{e+} > \omega_{e-}$, then value $(\mu_+^{pl}/\mu_+)/(\mu_-^{pl}/\mu_-)$ can be both
larger and less than unity (under big $\omega$) depending on the angular position of the source $y$. In
Figures 2,3 we see both situations, the solid lines)

b) if the rays forming the secondary image go through greater density than the rays forming the primary
image, that is $\omega_{e+} < \omega_{e-}$, then the value $(\mu_+^{pl}/\mu_+)/(\mu_-^{pl}/\mu_-)$ can be
only less than unity (see Figures 2,3, the dashed lines).

Critical value of $y$, separating two cases of the location of the solid line (below unity and above unity)
equals to
\begin{equation}
y_{crit} = \frac{\sqrt{-8 + 6 \sqrt{2}}}{2} \cong 0.348
\end{equation}
for $\omega_{e+}^2 = 2 \omega_{e-}^2$ (the solid lines on Fig.2), and
\begin{equation}
y_{crit} = \frac{\sqrt{-200 + 110 \sqrt{10}}}{10} \cong 1.216
\end{equation}
for $\omega_{e+}^2 = 10 \omega_{e-}^2$ (the solid lines on Fig.3).

\section*{Appendix 2. Calculation of the integral $Int2$.}

By substituting $1/\sqrt{z^2+b^2}=x$ the integral $Int2$ is reduced to
\begin{equation}
Int2 = \frac{1}{b} \int \limits_0^{1/b} \frac{x}{\sqrt{\frac{1}{b^2}
- x^2}} \, e^{Bx} \, dx .
\end{equation}
Integration can be performed with using of \citet{Gr-R}:
\begin{equation}
Int2 = \frac{1}{b^2} \left( 1 + \frac{\pi}{2} \left[ I_1(B/b) +
L_1(B/b) \right] \right),
\end{equation}
where $I_1$ is the Bessel function of the first kind and $L_1$ is
the Struve function.

Under condition $B/b \ll 1$ which corresponds to $G M_0/b \ll \Re T$, the integral $Int2$ is reduced to
\begin{equation}
Int2 = \frac{1}{b^2} \left( 1 + \frac{\pi}{4} \frac{G M_0}{b \Re T}
\right) .
\end{equation}

\section*{Acknowledgments}
This work was partially supported by RFBR grants 08-02-00491 and 08-02-90106, the RAN Program 'Formation and
evolution of stars and galaxies' and Russian Federation President Grant for Support of Leading Scientific
Schools NSh-3458.2010.2. The work of OYuT was also partially supported by the Dynasty Foundation and Russian
Federation President Grant for Support of Young Scientists MK-8696.2010.2.

\label{lastpage}

\end{document}